# Towards Understanding the Impact of Crime in a Choice of a Route by a Bus Passenger

Daniel Sullivan[1], Carlos Caminha[1], Hygor Melo[2], Vasco Furtado[1],

[1]UNIFOR – University of Fortaleza Av. Washington Soares 1321, Fortaleza CE, Brazil
[2]IFCE – Instituto Federal de Educação, Ciência e Tecnologia do Ceará Av. Des. Armando de Sales Louzada, Acaraú CE, Brazil

<daniel.sullivan@edu.unifor.br>, {caminha, vasco}@unifor.br, <hygor@fisica.ufc.br>

**Abstract.** In this paper we describe a simulation platform that supports studies on the impact of crime on urban mobility. We present an example of how this can be achieved by seeking to understand the effect, on the transport system, if users of this system decide to choose optimal routes of time between origins and destinations that normally follow. Based on real data from a large Brazilian metropolis, we found that the percentage of users who follow this policy is small. Most prefer to follow less efficient routes by making bus exchanges at terminals. This can be understood as an indication that the users of the transport system privilege the security factor.

## 1   Introduction

The academic community has reported the correlation between human mobility and crime from different perspectives. Examples of this are the works by Cohen and Felson [11] that relates the occurrence of crimes by the convergence of routines between offenders and unprotected victims, Brantingham and Brantingham [35] works on environmental criminology and, most recently, Caminha et al. [17], who found that the relationship between clusters of floating population and crimes against property within a large metropolis follows a power law.

Despite the fact that the correlation between mobility and crime is widely recognized and studied, few quantitative studies have been developed to understand the impact of crime on the mobility of people in large cities. However, digitized data on the movement of people are increasingly abundant, opening up the possibility of elaborating more complete social models that can be validated.

Our research in this context focuses on data on the movement of people in buses of a large Brazilian metropolis, the city of Fortaleza. These are data on about one million people who daily use buses on their journeys, which characterizes this mode of transportation as the most important form of city displacement (the total population of the city is 2.4 million people).

In previous studies, we have studied the impact of the movement of people on the occurrence of crime [17] as well as on police allocation strategies [36]. In this article we will follow a different strategy, as we will investigate the impact that crimes distributed in the city can have on the choice of bus routes made by people. We have developed a software platform that allows, from data of movement and crime in the city, to simulate the impact on the public transportation system. In particular, we simulate the negative effects that this can bring on the behavior of bus users leading them to choose clearly less efficient routes in terms of travel time, comfort and / or distance traveled. In addition, it was possible to estimate the impact on the public bus transportation system in general by evaluating indicators such as the passenger's waiting time in bus stops and bus overcrowding.

The methodology that we followed was to compare two models of bus routes. In the first moment, based on the real world data on the track of users, we evaluate the system when the users make their actual routes. Then we investigated how the Public Transport System (PTS) would behave if users chose to make more efficient routes in terms of time. In order to achieve this goal, users are likely to seek to make bus line changes in order to minimize route time since many origin-destination routes are impossible to complete from a single direct route. This strategy, privileging the time of the journey, is not necessarily the most efficient in terms of security, since bus exchange places can be crime hot spots. In fact, It has been found that, in Fortaleza, the bus stops the most appropriate to people commute are significantly more insecure than the places that users actually do. This result raises the possibility that users of the Fortaleza bus system may be making more inefficient routes to escape crime.

## 2   Related Work

Agent-based modeling provides a simplified simulation of the reality, but is also a powerful technique to replicate social phenomena. The agent-based model uses a bottom up approach simulating the individual behavior of multiple agents in order to predict complex phenomena. These general characteristics allow a large number of applications in a diverse range of areas, such as:
archaeology [1,2]; biological models of infectious diseases [3]; growth of bacterial colonies [4]; alliance formation of nations during the Second World War [5]; modelling economic processes as dynamic systems of interacting agents [6]; size-frequency distributions for traffic jams [7].

In this work our main interest are in simulations of crime and urban mobility dynamics.

In Criminology, the role of urban space and its social relations has been previously emphasized to explain the origin of Crime [8, 9]. Particularly, the routine activity theory, proposed by Cohen and Felson [8], states that crimes, more specifically property crimes, such as robbery and theft, occur by the convergence of the routines of an offender, motivated to commit a crime, and an unprotected victim. The dynamics of crime and the impact of social relations on the increase of violence has been the object of study in several areas such as Social Sciences [10], Criminology [8,9,11], Computing [12-17], Economics [18] and Physics [19-25]. There are also a number of papers that use simulations in Criminology in order to test theories [26]; to study burglary including transportation network and statistical based human mobility patterns along the network [27]; to analyze police patrol routes [28].

There is a vast literature on human mobility [29-33]. However, less attention are given to the possible connection with crime. Some new researches show that human mobility is a key factor to understand the spatial distribution of crime in the urban landscape [17, 34]. Yet, works how crime can affect the human mobility are scarce.

## 3  Data Sets

We used user routes and a bus travel time network produced in a Data Mining project executed on mobility data in the city of Fortaleza [17,31,37]. A user's route on a bus network is defined by bus stops visited in sequence by that user within one or more buses. Formally, a user route $R_u$,, is a set containing $n$ ($\{p_1, p_2, p_3, \ldots ,p_n\}$) bus stops visited in sequence by a set of bus lines, $L_u = \{l_1, l_2, l_3, \ldots ,l_m$, used by $u$ in its course. Furtado et al [37] estimated these routes from the actual origin of each user to the actual destination, see [31] for details. For each origin and destination, two routes were estimated; one that the authors consider to be the actual route of the user and the other is an optimal route in time. In total, 294870 origins and destinations (representing an average behavior for business days in Fortaleza) were estimated, starting from these origins to their respective destinations, the same number of real and optimal routes was estimated.

Access was obtained to the scheduled route of the city buses, for each bus line if there is the quantity of vehicles available on a weekday, capacity of each vehicle, departure time for each trip of each vehicle and the expected arrival time.

Also from [37], a network of time was obtained, where the nodes represent bus stops and the links the average time between these two bus stops. them. Each link was assigned a probability function that is based on the mean time ($\mu$) and standard deviation (σ) calculated from the actual GPS data of the buses. Formally, the time of an edge at time x is obtained by the function:

$$f(x,\mu,\sigma) = \frac{1}{\sqrt{2\pi\sigma^2}} e^{\left(\frac{-(x-\mu)^2}{2\sigma^2}\right)}, -\infty < x < \infty, \sigma > 0 \qquad (1)$$

Crime data refers to crime against property (theft, robbery and burglary) occurred between 9/23/2014 and 10/4/2016 in the city of Fortaleza and metropolitan region. In total, this dataset has 98431 crimes. Figure 1 shows the density map of these crimes in Fortaleza.

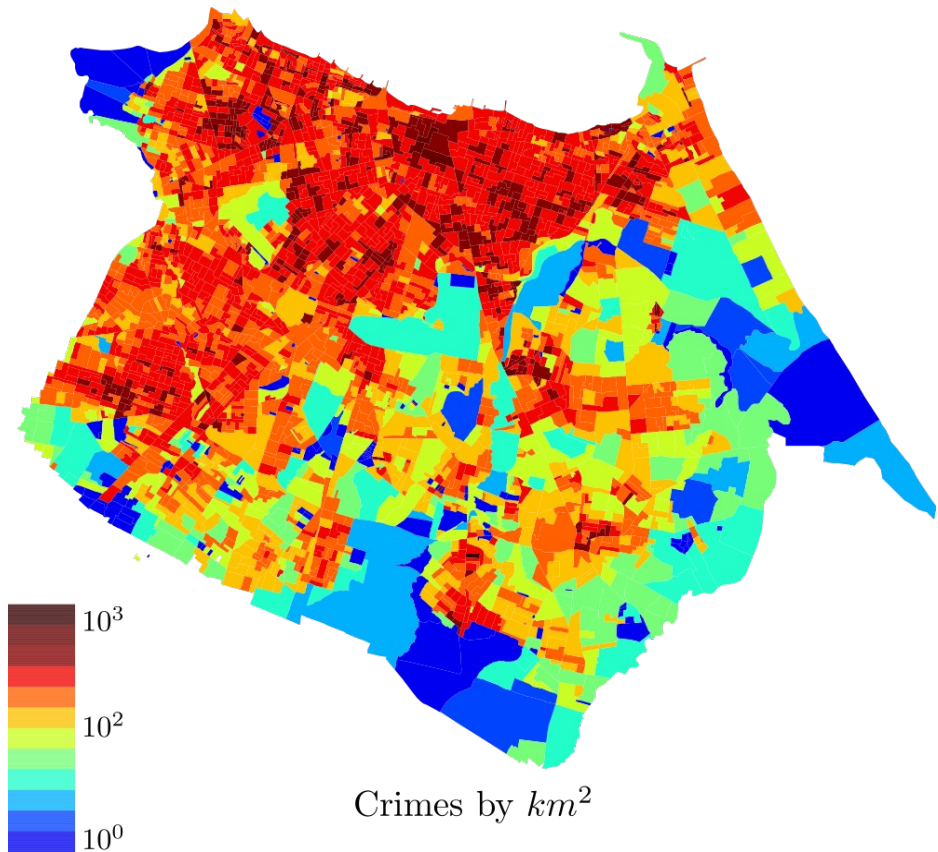

Fig. 1. Density map for property crime in Fortaleza from 2014 to 2016.

# 3  Simulation

The simulator was developed using Repast, a java multi-agent simulation framework. This simulator handles events whose management and execution of actions is done by a clock, which has a tick as a measure of time. The tick is incremented when all scheduled actions are performed.

The main elements of this tool are: Bus; Line; Route; Passenger; Stops; and Crimes. Buses behave like agents in the simulation, their behavior is stopped by the stops and make trips exchanges. They have as main attributes the current stop (corresponds to the stop it is or the last stop it has visited), the next stop to be visited, the time needed to reach the next stop, the list of passengers in the vehicle, the Current line and a hash that commits the trips that the bus will make in one day. Passengers, in turn, have information regarding the route they will take, ie they have a list with the lines that they will use and another one regarding the stops that they will board and disembark. Passengers are agents who board and disembark from buses when vehicles pass through their initial and destination stop. The lines have a list with all stops on your route. The crimes have information on where they occurred (latitude and longitude) and their type.

It was defined that the simulation start time will be two o'clock in the morning, because it is the time when there are fewer buses in circulation and fewer passengers using the system. This minimizes the effect of simulation start with empty buses. The simulation closes at 2:00 p.m., and it is possible to evaluate the system in its peak hours, from 5:00 am to 8:00 am, as well as at low usage times, for example at dawn. Every minute is called a bus dispatch event, in this event it is checked if the time of departure of each bus has already been reached, if positive the bus enters the simulation, at the end of this event is increased one minute in the time of the simulator.

Each bus has an event called move, it consists of all the logic that makes the buses walk in the network. This event is triggered after the dispatch event. After all buses run their event, the process will be finalized and will only start again at the next tick. As already mentioned a bus has a current stop and the next stop, the time in which the bus will take to reach the next stop will be calculated using the average time in which the buses of this line take to go through this stretch and the standard deviation of those times , These are the input data to apply the normal or Gaussian distribution and get the time in which the bus will take to go along an edge.

Each tick represents one minute in the time of the simulator, if the time for the bus to cross the edge is greater than sixty seconds, the bus will not be able to complete this stretch still in this tick, so the time for it to arrive at the next stop will only be decremented from sixty Seconds. If the time is exactly equal to sixty seconds, it will travel this edge still in this movement and will arrive at the next stop. The last case that can occur is the time being less than sixty seconds, that is, one bus will arrive at the next stop and will continue towards another.

Whenever a bus arrives at a stop, it is checked if there are passengers intending to disembark or board the vehicle. A bus in Fortaleza has 80 passengers, so if its capacity is reached and some passenger wants to embark, it will have to wait for the next vehicle of that line to pass.

When the bus of a certain line reaches the end of its route, it is checked if there are still trips for that vehicle to carry out, if there is a check that the line is maintained or if there is a line change, the trip changes are made and the bus Start a new trip. When there are no more trips for that bus to perform, it will be removed from the simulation. The simulator closes at 2 o'clock in the morning the next day or when all buses finish all their trips.

## 4  Methodology and Empirical Evaluation

In this article, the bus system will be evaluated by waiting time for shipment, stocking of vehicles and safety at the point of transfer. The average waiting time for boarding is calculated by subtracting the time of boarding less the time that the passenger appeared at the stop, and the vehicle stocking distribution, which shows the quantities of vehicles with low occupancy (between 0 and 20 passengers ), Intermediate occupation (between 21 and 60 passengers) and high occupancy (between 61 and 80). Security was measured by the ratio of the number of crimes occurring in the vicinity of the transfer point by the number of users who are on-site, we call this Rate of Crime.

The behavior of the network in its natural operation was observed, i.e. when all users make their actual route. The behavior of the network was also evaluated in a hypothetical scenario, simulating a situation where all users make optimal paths of time. The contrasts between the values of the individual indicators adopted in this research can be observed in the panel of Figure 2.

Figure 2 (A) shows the stocking of vehicles when all users make actual routes, while in Figure 2 (B) the vehicles are stocked for the optimal routes. It may be noted that when passengers make optimal routes there are less crowded buses. This is because Fortaleza's transit network is not prepared for its users to make optimal routes in time, thus, many passengers are waiting for vacancies to appear on the few crowded buses. In Figure 2 (C) it can be seen that at the beginning of the day, where there are few buses running, the waiting time for the passengers is similar, regardless of the route (real or optimal). However, from the peak time onwards, the number of vehicles in circulation increases and the waiting time for those who make the optimal routes becomes longer than the waiting time for actual routes. This is because when everyone

does the right thing, there are not enough vehicles to meet the demand. In Figure 2 (D) it is possible to visualize the crime rate per stop at optimal and real ballooning points. It is observed that the risk at optimal points of transfer is greater than at the points where users actually pay, especially outside the peak hours (from 5 am to 8 am), possibly the largest number of people on the street discourages criminals from committing Crimes at that time. This result raises the possibility that people choose slower routes to escape the crime, essentially because the slower route has its point of transfer in a bus terminal, a place that offers, among other things, greater safety for users.

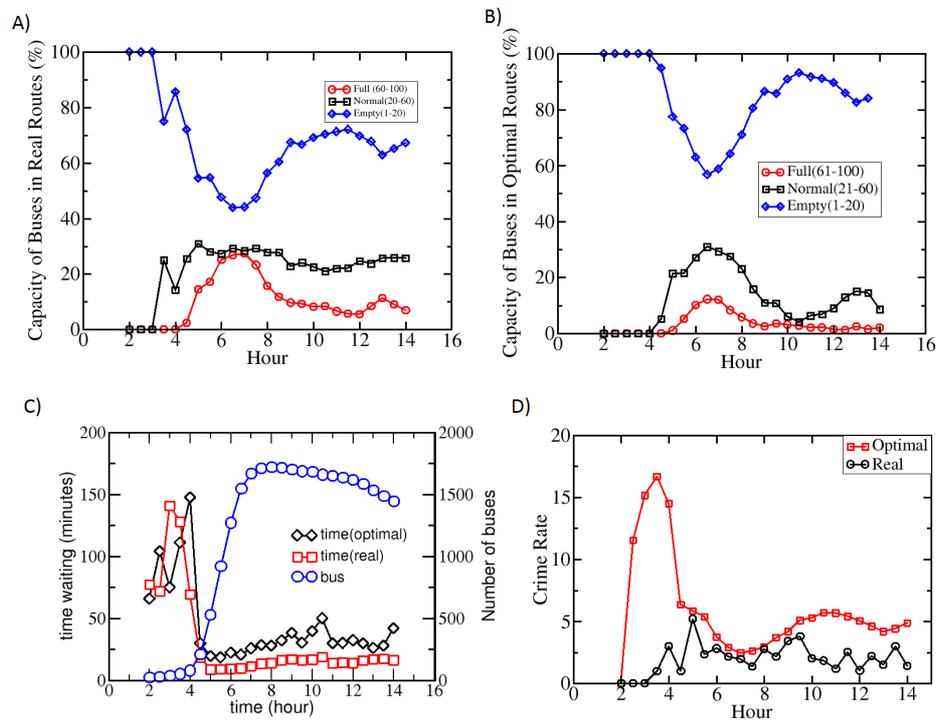

Fig. 2. Indicators of quality of use and safety for the bus network of Fortaleza.

In Fig. 3 we show the relation between crimes and the urban public transportation network. Recently, an extensive analysis conducted over real crime data related to a large Brazilian metropolis [38] demonstrated that the spatial distribution of crimes such as robberies, thefts, and burglaries follows a power law, more specifically, a Zipfian distribution [39]. We see in Fig. 3A that the distribution of crimes occurring close of bus stops visually follows a power law, with exponent close of -2.0 as show by the guide red line. The power law distribution indicates that we can find bus stops with much large number of crimes when we compare with an exponential distribution. In order to see the effect of crimes on the utilization of the urban transportation we computed the number of crimes that occurred close of a bus stop that can be utilized for a connection transfer at the optimal routes. As we already mentioned a large fraction of the users do not use the optimal connection transfer, instead of this, they seem to choose a safe route. In Fig. 3B we show that the large fraction of connexions transfer in the optimal route (98\%) have more than 10 crimes close. This is an indication that the users prefer to go by a safe but large trajectory.

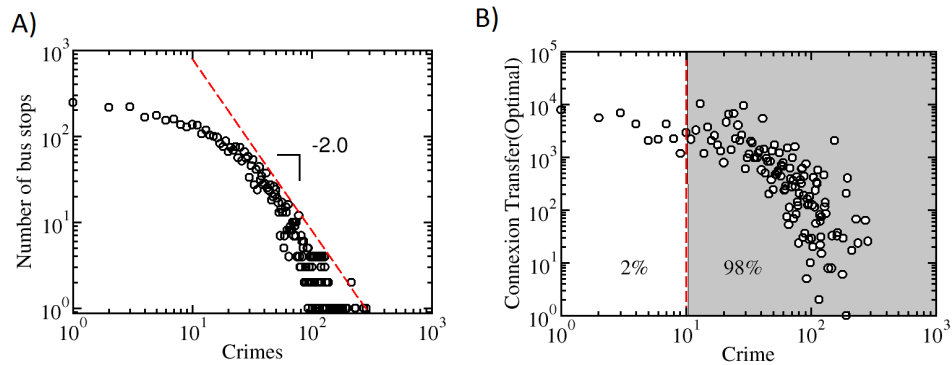

Fig. 3. Histograms of crimes.

## 5 Conclusion

In this paper we describe a simulation platform that supports studies on the impact of crime on urban mobility. We present an example of how this can be achieved by seeking to understand the effect, on the public transport system, if riders decide to choose optimal routes of time between origins and destinations. Based on real data from a large Brazilian metropolis, we found that, when exchange of bus is necessary, the percentage of users who needs this policy is small. Most prefer to follow less efficient routes by making bus exchanges at terminals. This can be understood as an indication that the users of the transport system privilege the security factor. This indication is corroborated by the fact that 98% of the optimal routes going through unsafe bus stops (more than 10 crimes).

Our research is ongoing and new tests are necessary to detail the features the riders consider the most. The simulation environment must also be improved to help this, but this first essay has already shown to us that it is an important to tool to help understanding the impact crime on urban mobility. Based on real data describing the geographic space, we intend to further investigate particularly new different policies based on features like comfort an overcrowding in order to tune the simulation.